\documentclass[conference]{IEEEtran}
\usepackage{multirow}
\usepackage{amssymb}
\usepackage{amsmath}
\usepackage{mathrsfs}
\usepackage{amsfonts}
\newtheorem{*theorem}{Theorem}
\usepackage{graphicx}
\usepackage{color}
\usepackage{cite}
\usepackage{amsmath}
\usepackage{extarrows}

\newtheorem{lemma}{\bf{Lemma}}[section]
\newtheorem{remark}{\bf{Remark}}

\newcommand{\bx}{\mbox{\boldmath $x$}}

\newcommand{\bbf}{\mbox{\boldmath $f$}}
\newcommand{\baf}{\mbox{\boldmath $\bar{f}$}}
\newcommand{\bA}{\mbox{\boldmath $A$}}
\newcommand{\bB}{\mbox{\boldmath $B$}}
\newcommand{\bG}{\mbox{\boldmath $G$}}

\newcommand{\bJ}{\mbox{\boldmath $J$}}
\newcommand{\bK}{\mbox{\boldmath $K$}}

\newcommand{\bR}{\mbox{\boldmath $R$}}
\newcommand{\bC}{\mbox{\boldmath $C$}}
\newcommand{\bE}{\mbox{\boldmath $E$}}
\newcommand{\bF}{\mbox{\boldmath $F$}}
\newcommand{\bI}{\mbox{\boldmath $I$}}
\newcommand{\bX}{\mbox{\boldmath $X$}}
\newcommand{\bT}{\mbox{\boldmath $T$}}
\newcommand{\bY}{\mbox{\boldmath $Y$}}

\newcommand{\bZ}{\mbox{\boldmath $Z$}}
\newcommand{\by}{\mbox{\boldmath $y$}}
\newcommand{\bh}{\mbox{\boldmath $h$}}
\newcommand{\bn}{\mbox{\boldmath $n$}}
\newcommand{\bzero}{\mbox{\boldmath $0$}}

\newcommand{\bz}{\mbox{\boldmath $z$}}

\newcommand{\Tr}{\textnormal{Tr}}

\newcommand{\HH}{\dagger}

\newcommand{\bUps}{\mbox{\boldmath $\Upsilon$}}
\newcommand{\bups}{\mbox{\boldmath $\upsilon$}}

\begin{document}
\bibliographystyle{IEEEtran}
\title{Joint Power Splitting and Secure Beamforming Design in the Wireless-powered Untrusted Relay Networks}

\author{\IEEEauthorblockN{Mingxiong Zhao$^{\HH}$, Suili Feng$^{\HH}$, Yuan Liu$^{\HH}$, Xiangfeng Wang$^{*}$, Meng Zhang$^{\HH}$, Hao Fu$^{\HH}$ \thanks{\IEEEauthorrefmark{4} This work is supported by the the National Natural Science Foundation of China under grants 61340035 and 61401159, the Science $\&$ Technology Program of Guangzhou under grant 2014J4100246, the SCUT-UNSW Canberra Research Collaboration Scheme, and the Chinese Scholarship Council under
Grant 201206150017.}}
\IEEEauthorblockA{$^{\HH}$School of Electronic and Information Engineering, South China University of Technology,Guangzhou, China\\
$^{*}${Shanghai Key Lab. of Trustworthy Computing, Software Engineering Institute, East China Normal Univ., Shanghai, China}\\
Email: jimmyzmx@gmail.com, \{fengsl,eeyliu\}@scut.edu.cn, xfwang@sei.ecnu.edu.cn}}

\maketitle
\begin{abstract}
In this work, we maximize the secrecy rate of the wireless-powered untrusted relay network by jointly designing power splitting (PS) ratio and relay beamforming with the proposed global optimal algorithm (GOA) and local optimal algorithm (LOA). Different from the literature, artificial noise (AN) sent by the destination not only degrades the channel condition of the eavesdropper to improve the secrecy rate, but also becomes a new source of energy powering the untrusted relay based on PS. Hence, it is of high economic benefits and efficiency to take advantage of AN compared with the literature. Simulation results show that LOA can achieve satisfactory secrecy rate performance compared with that of GOA, but with less computation time.
\end{abstract}

\section{Introduction}
Recently, simultaneous wireless information and power transfer (SWIPT) became an emerging solution for prolonging the lifetime of energy-constrained wireless nodes and draw significant attention in the cooperative transmission. Since SWIPT enables receivers to harvest energy and encode information from the same wireless signal, it makes the most efficient utilization of wireless spectrum for both information and energy transfer. SWIPT has drawn a great deal of research interests \cite{ZhangHo,Zhou,Liuz,LiuPS}. Two practical schemes, namely power splitting (PS) and time switching (TS), were proposed in \cite{ZhangHo} and \cite{Zhou}. For TS, the receivers switch over time between information decoding and energy harvesting. PS enables wireless nodes to split some power for information decoding and use the remaining power for energy harvesting. On the other hand, energy harvesting powered relay systems have gained significant interests, where the relays with SWIPT use their harvested energy to power the forwarding of the sources' information without external power supplies.
In \cite{EHrel}, time switching based relaying (TSR) protocol and power splitting based relaying (PSR) protocol were proposed for wireless-powered relays systems. The authors in \cite{VolEH} investigated energy harvesting relays networks where relays switched over time between harvesting energy and volunteering to forward the transmitter's information. In \cite{PSAS}, a multiple-antenna relay system  with SWIPT was studied and a ``harvest-and-forward" strategy was proposed to maximize the achievable rate. The authors of \cite{two-wayEH} investigated the beamforming design with simultaneous energy harvesting to improve the max-min signal-to-interference-plus-noise ratio (SINR) in a multiuser two-way multi-antenna relay systems.

On the other hand, with the development of wireless networks, the issues of privacy and security have attracted much interest and attention due to the broadcast nature of the wireless medium. The cryptographic approach using secret keys at upper layers has been traditionally used for the security of communication systems. Due to the cost of the key distribution and key management, information-theoretic security at the physical layer became an important complementary to the traditional cryptographic approach. Thus, substantial research efforts have been dedicated to information-theoretic physical layer security\cite{Shannon,Wyner,Coop,Untrust,MIMOuntrusted,Goel,twountr,zhao2014secure,li2013spatially,liu2013destination}, the concept of which was defined by Wyner \cite{Wyner}. By exploiting spatial diversity, a relay may also provide the secrecy capacity by assisting the source-destination transmission or acting as a jammer (e.g. \cite{Coop}). However, another security threat arises since the information transmitted by the source may be eavesdropped by an untrusted relay \cite{Untrust}. Interestingly, in \cite{Untrust}, it was shown that the secrecy rate may be enhanced even with the help of an untrusted relay. In \cite{MIMOuntrusted}, the authors studied joint secure beamforming for an AF untrusted relay system. The MIMO AF untrusted relay beamforming was further extended to a two-way untrusted relay system in \cite{twountr}.
To further improve the security of the source-destination transmission, artificial noise (AN) (sent by jammer\cite{Coop}, source\cite{li2013spatially} or destination\cite{liu2013destination}) was proposed in \cite{Goel} as another important solution for physical layer security. Traditionally, AN is only used to degrade the channel conditions of eavesdroppers, hence, it may not be efficient and economic from the aspect of green communication.

More recently, a handful of works studied the SWIPT with the consideration of physical layer security \cite{LiuGC13,NgGC13,li2014secure,EHjammer}. The authors in \cite{li2014secure} studied secure relay beamforming for SWIPT in one-way relay systems with an external energy-harvested receiver. In \cite{EHjammer}, the authors considered a wireless-powered jammer system, where the jammer can harvest energy from the wireless signal and use it to interfere with the eavesdropper. In the present literature, the energy of AF relay, especially the untrusted relay, is considered mostly to be supplied by the power grid or battery. To the best of our knowledge, secrecy issues in the wireless-powered untrusted relay systems have not been studied yet, which motivates this work. And it seems reasonable and intuitive to realize the energy harvest at the untrusted AF relay node to make full use of AN.

In this work, we consider a wireless-power untrusted relay network and maximize its secrecy rate by jointly designing PS ratio and relay beamforming with the proposed global optimal algorithm (GOA) and local optimal algorithm (LOA). The main difference from the literature is that AN sent by the destination not only degrades the channel condition of the eavesdropper to improve the secrecy rate, but also becomes a new source of energy powering the PS based untrusted relay. Hence, it is of high economic benefits and efficiency to take advantage of AN compared with the literature.

\emph{Notation:} We adopt the notation of using boldface for column vectors (lower case), and matrices (upper case). The hermitian transpose is denoted by the symbol $(\cdot)^\HH$. For a complex scalar $x$, its complex conjugate is denoted by $x^*$. $E[\cdot]$ and $\cal{CN(\cdot)}$ denote the statistical expectation and complex Gaussian distributions, respectively. For matrix $\bX$, vec($\bX$), $\bX^{-1}$, $\bX^T$, and Tr($\bX$) represent the vectorization, inversion, transpose and trace of matrix $\bX$, respectively, and $\bX\otimes\bY$ stands for the kronecker product of $\bX$ and $\bY$. For a vector $\bx$, we use $\|\bx\|_2$ to indicate its $\ell_2$ norm. And $\bI$ is the identity matrix with corresponding dimensions.

\section{System model}
Consider a scenario with three nodes: an untrusted AF relay (R), source S and destination D, where only R is equipped with multiple antennas, and the scenario with single-antenna R is regarded as a special case in this paper. Due to the space limitation, the multi-antenna case where all the nodes own multiple antennas is considered in the journal version.

\begin{figure}[t]
  \centering
  \includegraphics[height=45mm]{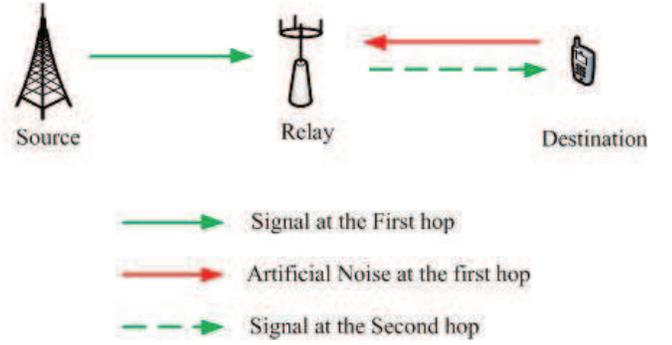}\\
    \caption{Simulation Scenario.}\label{1}
\end{figure}

In this paper, the considered relay R is untrusted and assumed to eavesdrop the confidential signal of D when it helps the SD transmission, and it can be treated as a legitimate user \cite{Untrust,MIMOuntrusted,Goel,twountr} with different service from that of D. In Fig.\ref{1}, at the first hop, S transmits the confidential signal to R, meanwhile, D sends AN to R to degrade the eavesdropping channel of R. In contrast to the literature, R can not only process the received signal, but also harvest energy from it as the source of power for the next-hop transmission at the same time. At the second hop, R transmits the processed signal to D depending on the harvested energy. Notice that both R and D are half-duplex, i.e., R and D can not receive and send signal, simultaneously. Therefore, SD link is not under our consideration.

Denote the channel vectors from S to R, from D to R, and from R to D by $\bh_{sr}\in \mathbb{C}^{N_{r} \times 1}$, $\bh_{dr}\in \mathbb{C}^{N_{r} \times 1}$ and $\bh_{rd}\in \mathbb{C}^{N_{r} \times 1}$, where $N_{r}$ is the number of antennas at R. At the first hop, S transmits the signal to R, meanwhile D sends AN to R with the purpose of confounding the eavesdropper. The received signal at R is given by
\begin{equation}
\by_r=\bh_{sr}x_s+\bh_{dr}x_d+\bn_r,
\end{equation}
where $\bn_r \in \mathbb{C}^{N_r}$ represents the additive white Gaussian noise (AWGN) associated with relay node following the distribution ${\cal{CN}}(\bzero,\sigma^2_r\bI)$, $x_s$ represents the secure information sent by S to D, meanwhile, $x_d$ is AN sent by D to degrade the channel condition of R. And $x_s\in\mathbb{C}$ and $x_d\in\mathbb{C}$ represent the transmit symbol of S and D with powers $E[x_sx_s^*]=P_s$ and $E[x_dx_d^*]=P_d$. $\rho$ denotes the ratio of power split for the energy harvesting, and the information processing is $(1-\rho)$ at the relay node. Hence, the parts of received signal for energy harvesting and information processing are written, respectively, as follows
\begin{eqnarray}
\tilde{\by_r}&=&\sqrt{\rho}\left(\bh_{sr}x_s+\bh_{dr}x_d+\bn_r\right),\label{equ:part for EH}\\
\breve{\by_r}&=&\sqrt{1-\rho}\left(\bh_{sr}x_s+\bh_{dr}x_d+\bn_r\right)+\bn_c.\label{equ:part for IP}
\end{eqnarray}
where $\bn_c \in \mathbb{C}^{N_r}$ is the additive noise vector introduced by signal conversion from the passband to baseband with the distribution ${\cal{CN}}(\bzero,\sigma^2_c\bI)$.

Based on \cite{Ar99capacityof}, we denote $\bA(\rho):=(1-\rho)P_d\bh_{dr}\bh_{dr}^{\HH}+(1-\rho)\sigma_r^2\bI+\sigma_c^2\bI$. Furthermore, we can rewrite (\ref{equ:part for IP}) as
\begin{eqnarray}
\label{equ1:part for EH}
\hat{\by_r}\!&=&\!\sqrt{1-\rho}\bA^{-\frac{1}{2}}(\rho)\bh_{sr}x_s+\nonumber\\
&&\bA^{-\frac{1}{2}}(\rho)\left(\sqrt{1-\rho}\bh_{dr}x_d+\sqrt{1-\rho}\bn_r+\bn_c\right),
\end{eqnarray}
and the information rate achieved at the untrusted relay can be given by
\begin{equation}
\label{equ:information rate of relay}
R_r(\rho)=\text{log}_2\left(1+(1-\rho)P_s\bh_{sr}^{\HH}\bA^{-1}(\rho)\bh_{sr}\right).
\end{equation}

Resorting to the PS scheme, we can briefly write the expression of harvested energy at R as
\begin{equation}
\label{equ:expression of EH}
P_r^{\text{EH}}=\eta\left\{\rho P_s\left\|\bh_{sr}\right\|_2^2+\rho P_d\left\|\bh_{dr}\right\|_2^2+\rho \sigma_r^2\Tr\left(\bI\right)\right\},
\end{equation}
where $\eta\in (0,1]$ denotes the energy conversion efficiency from signal power to circuit power.

Different from simply applying the amplified scalar $\alpha$ to tackle the received signal at the single-antenna R in the Remark \ref{remark1}, in this paper, we adopt $\bF$ as the processing matrix to deal with the received signal and forward it to D at relay node. Subtracting the AN term $\sqrt{1-\rho}\bh_{rd}^{\HH}\bF \bh_{dr}x_d$ sent by itself, the received signal at D can be given by
\begin{eqnarray}
\label{equ1:received signal at D}
y_d=\bh_{rd}^{\HH}\bF\left(\sqrt{1-\rho}\bh_{sr}x_s+\sqrt{1-\rho}\bn_r+\bn_c\right)+n_d,
\end{eqnarray}
where $n_d$ is the AGWN introduced by the receiver antenna at D with the distribution ${\cal{CN}}(0,\sigma_d^2)$. For simplicity of description, we make $\sigma_r^2=\sigma_c^2=\sigma_d^2=\sigma^2$ in the following parts. The achievable rate of D can be expressed as
\begin{equation}
\label{equ:information rate of D}
\!R_d(\bF\!,\!\rho\!)\!=\!\text{log}_2\!\!\!\left(\!\!\!1\!\!+\!\!\frac{(1-\rho)P_s\left\|\bh_{rd}^{\HH}\bF\bh_{sr}\right\|_2^2}{\!(\!1\!-\!\rho\!)\sigma_r^2\left\|\!\bF^{\HH}\bh_{rd}\!\right\|_2^2\!\!\!+\!\sigma_c^2\left\|\bF^{\HH}\bh_{rd}\right\|_2^2\!\!\!+\!\!\sigma_d^2}\!\right)\!\!.
\end{equation}

\section{Problem Formulation For Secrecy Rate Maximization}
To begin with our problem formulation, we will provide a brief background on the introduction of \emph{Secrecy Rate}\cite{Coop,Untrust,MIMOuntrusted,Goel,twountr,zhao2014secure,li2013spatially,liu2013destination}. In the presence of the eavesdropper, the secrecy rate is the information rate that the source can securely transmit to the legitimate user. In this section, we denote $R_{sr}(\bF,\rho)$ as the secrecy rate, given by
\begin{equation}
\label{equ:secrey rate}
R_{sr}(\bF,\rho)=\frac{1}{2}\left[R_d(\bF,\rho)-R_r(\bF,\rho)\right]^+,
\end{equation}
where $[a]^+=\max(0,a)$, and $\frac{1}{2}$ is from the fact that the source-destination transmission takes place in two time slots.

For the MIMO relay case, the optimization problem can be written as
\begin{eqnarray}
\label{eqn:optimization problem}
\max_{\rho, \bF}\!\!\!\!\!\! &&R_{sr}(\bF,\rho)\\
\text{s.t.} \!\!\!\!\!\!&&\!(1\!-\!\rho)P_s\left\|\bF\bh_{sr}\right\|_2^2\!+\!(1-\rho)P_d\!\left\|\bF\bh_{dr}\!\right\|_2^2\!+\nonumber\\
&&\!(1-\rho)\sigma_r^2\left\|\bF\right\|_F^2\!+\!\sigma_c^2\left\|\bF\right\|_F^2\!\leq \! P_r^{\text{EH}},\nonumber
\end{eqnarray}
which can be briefly rewritten as
\begin{eqnarray}
\label{eqn1:optimization problem}
\max_{\rho, \bbf} &&\scriptsize{\left(\!1\!+\!\frac{\bbf^{\HH}\bB(\rho)\bbf}{\bbf^{\HH}\bC(\rho)\bbf+\bbf^{\HH}\bE\bbf+\sigma^2}\!\right)\!\!\left(\!\frac{1}{1+(1-\rho)P_s\bh_{sr}^{\HH}\bA^{-1}(\rho)\bh_{sr}}\!\!\right)}\nonumber\\
\text{s.t.} &&\bbf^{\HH}\bT(\rho)\bbf\leq P_r^{\text{EH}},
\end{eqnarray}

where
\begin{eqnarray}
\bbf&=&\text{vec}(\bF),\nonumber\\
\bB(\rho)&=&P_s(1-\rho)\left[\left(\bh_{sr}\bh_{sr}^{\HH}\right)^T\otimes\left(\bh_{rd}\bh_{rd}^{\HH}\right)\right],\nonumber\\
\bC(\rho)&=&(1-\rho)\left[\left(\sigma^2\bI\otimes\bh_{rd}\bh_{rd}^{\HH}\right)\right],\nonumber\\
\bE&=&\sigma_c^2\left[\left(\bI\otimes\bh_{rd}\bh_{rd}^{\HH}\right)\right],\nonumber\\
\bG(\rho)&=&(1-\rho)P_s\left[\left(\bh_{sr}\bh_{sr}^{\HH}\right)^T\otimes\bI\right],\nonumber\\
\bJ(\rho)&=&(1-\rho)P_d\left[\left(\bh_{dr}\bh_{dr}^{\HH}\right)^T\otimes\bI\right],\nonumber\\
\bT(\rho)&=&\left(\bG(\rho)+\bJ(\rho)+(1-\rho)\sigma^2\bI_f+\sigma_c^2\bI_f\right),\nonumber
\end{eqnarray}
where $\bI_f\in \mathbb{C}^{N_{r}^2 \times {N_{r}^2}}$ has the same dimension as $\bG(\rho)$, and the properties, such as $\text{vec}(\bA\bX\bB)=(\bB^T\otimes\bA)\text{vec}(\bX)$ and $\Tr{(\bX_1^T\bX_2)}=\text{vec}(\bX_1)^T\text{vec}(\bX_2)$, are applied to the above equalities. Note that the energy constraint in \eqref{eqn:optimization problem} means that the considered relay is wireless-powered, i.e., the relay is powered by its harvested energy without any external energy supplies.

\begin{remark}\label{remark1}
When S, D and R are equipped with the single antenna, the achievable rate of R and D are given by
\begin{eqnarray}
R_r(\rho)\!\!\!\!&=&\!\!\!\!\text{log}_2\left(1+\frac{(1-\rho)P_s|h_{sr}|^2}{(1-\rho)P_d|h_{dr}|^2+(1-\rho)\sigma_d^2+\sigma_c^2}\right),\nonumber\\
R_d(\rho)\!\!\!\!&=&\!\!\!\!\!\text{log}_2\left(\!\!1+\frac{\alpha^2(1-\rho)P_s|h_{rd}^{\HH}h_{sr}|^2}{\alpha^2(1-\rho)|h_{rd}|^2\sigma_r^2+\alpha^2|h_{rd}|^2\sigma_c^2+\sigma_d^2}\!\!\right),\nonumber
\end{eqnarray}
where $\alpha$ is the amplified factor with the consideration of power constraint at the relay, $\alpha=\sqrt{\frac{P_r^{\text{EH}}}{(1-\rho)(P_r|h_{sr}|^2+P_d|h_{dr}|^2+\sigma_r^2)+\sigma_c^2}}$. And the secrecy rate is
\begin{equation}
\label{equ:secrey rate in single case}
R_{sr}(\rho)=\frac{1}{2}\left[R_d(\rho)-R_r(\rho)\right]^+,
\end{equation}
which is a function of $\rho$ in the range of $(0,1)$, and can be perfect solved by setting $\frac{d R_{sr}}{d\rho}=0$ (resorting to Matlab), where $\frac{d R_{sr}}{d\rho}$ is the derivative of $R_{sr}(\rho)$ with respect to $\rho$. Note that there might be more than one point making the derivative $\frac{d R_{sr}}{d\rho}$ equals zero, and we treat the one yielding    the maximal secrecy rate $R_{sr}(\rho^\star)$ as the optimal PS ratio $\rho^\star$. If there is no point at which the derivative is zero, the secrecy rate is zero.
\end{remark}

\subsection{Global Optimal Solution}\label{GOS}
It is obvious that (\ref{eqn1:optimization problem}) is non-convex and difficult to solve with the two variables $(\rho,\bbf)$, simultaneously. In this subsection, we design the global optimal algorithm (GOA) for problem (\ref{eqn1:optimization problem}) with one dimension search of $\rho$ in the interval of $[0,1]$. In each round, the following optimization problem with the given $\bar{\rho}$ is considered to obtain the optimal solution $\bbf^\star$,

\begin{eqnarray}
\label{eqn1:optimization problem with given rho}
\max_{\bbf} &&\frac{\bbf^{\HH}\bB(\bar{\rho})\bbf}{\bbf^{\HH}\bC(\bar{\rho})\bbf+\bbf^{\HH}\bE\bbf+\sigma^2}\\
\text{s.t.} &&\bbf^{\HH}\bT(\bar{\rho})\bbf\leq P_r^{\text{EH}}.\nonumber
\end{eqnarray}
However, the above optimization problem is still a non-convex problem. Hence, we resort to semidefinite relaxation (SDR) to drop the rank-one constraint ($\text{rank}(\bX)=\text{rank}(\bbf\bbf^{\HH})=1$), and obtain the relaxed problem as follows
\begin{eqnarray}
\label{eqn:relaxed problem with given rho}
\max_{\bX} &&\frac{\Tr{\left(\bB(\bar{\rho})\bX\right)}}{\Tr{\left(\bC(\bar{\rho})\bX\right)}+\Tr{\left(\bE\bX\right)}+\sigma^2}\\
\text{s.t.} &&\Tr{\left(\bT(\bar{\rho})\bX\right)}\leq P_r^{\text{EH}}, \nonumber\\
&&\bX\succeq\bzero,\nonumber
\end{eqnarray}
which is a quasi-convex problem. Thus, we adopt the Charnes-Cooper transformation to make the quasi-convex problem (\ref{eqn:relaxed problem with given rho}) into a convex SDP. To this end, we set $\bX=\frac{\bZ}{\xi},\ \xi>0$, and recast (\ref{eqn:relaxed problem with given rho}) as
\begin{eqnarray}
\label{eqn1:relaxed subproblem with given rho}
\max_{\bZ,\xi} &&\Tr{\left(\bB(\bar{\rho})\bZ\right)}\\
\text{s.t.} &&\Tr{\left(\bC(\bar{\rho})\bZ\right)}+\Tr{\left(\bE\bZ\right)}+\xi\sigma^2=1,\nonumber \\
&&\Tr{\left(\bT(\bar{\rho})\bZ\right)}\leq \xi P_r^{\text{EH}}, \nonumber\\
&&\bZ\succeq\bzero,\xi > 0, \nonumber
\end{eqnarray}
the optimal solution of which can be efficiently obtained by solvers, e.g., CVX. Denote $(\bZ_{\bar{\rho}}^\star,\xi_{\bar{\rho}}^\star)$ as the optimal solution of problem (\ref{eqn1:relaxed subproblem with given rho}) with given $\bar{\rho}$, and we achieve the optimal solution of $\bX_{\bar{\rho}}^\star$ by $\bX_{\bar{\rho}}^\star=\frac{\bZ_{\bar{\rho}}^\star}{\xi_{\bar{\rho}}^\star}$. Note that the optimal $\bX$ with rank-one property, is the sufficient and necessary condition for the equivalence of problem (\ref{eqn1:optimization problem with given rho}) and (\ref{eqn:relaxed problem with given rho}), which is guaranteed by the following lemma:
\begin{lemma}\label{lemma_zhang}
\cite[Theorem 2.1]{huang2007complex} Suppose that $\bUps \in \mathcal{H}^n$ is a complex Hermitian positive semidefinite matrix of rank $r$, and $\bA_1$, $\bA_2\in \mathcal{H}^n$ be two given Hermitian positive semidefinite matrices. Then, there is a rank-one decomposition of $\bUps$.
\begin{equation}
\bUps=\sum\limits_{j=1}^r\bups_j\bups_j^{\HH}
\end{equation}
such that
\begin{equation}
\!\!\!\bups_j^{\HH}\bA_1\bups_j\!\!=\!\!\frac{\Tr{(\bA_1\bUps)}}{r}\!,\!\ \bups_j^{\HH}\bA_2\bups_j\!=\!\frac{\Tr{(\bA_2\bUps)}}{r},\!\ j\!=\!1,\cdots,r.
\end{equation}
\end{lemma}

In our problem, if we want to guarantee the rank-one property, the following equalities should be satisfied:
\begin{eqnarray}
\!\!\!\!\Tr{\left(\!\bC(\bar{\rho})\bZ\!\right)}\!\!+\!\!\Tr{\!\left(\bE\bZ\!\right)}\!\!+\!\!\xi\sigma^2\!\!\!\!\!\!\!&=&\!\!\!\!\!\!\!\Tr{\left(\!\!\bC(\bar{\rho})\bbf\bbf^{\HH}\!\right)}\!\!+\!\!\Tr{\left(\!\!\bE\bbf\bbf^{\HH}\!\!\right)}\!\!+\!\!\xi\sigma^2\!,\label{relaxed constraint 1}\\
\Tr{(\bT(\bar{\rho})\bX)}&=&\Tr{(\bR(\bar{\rho})\bbf\bbf^{\HH})}.\label{relaxed constraint 2}
\end{eqnarray}

Based on Lemma\ref{lemma_zhang}, we can recover the rank-one solution $\bbf_{\bar{\rho}}^\star$ from $\bX_{\bar{\rho}}^\star=\bbf_{\bar{\rho}}^\star\bbf_{\bar{\rho}}^{\HH\star}$ with given $\bar{\rho}$. For the sake of brevity, denote $f(\rho,\bbf)\triangleq f_1(\rho,\bbf)f_2[t(\rho)]$, where $f_1(\rho,\bbf)=1\!\!+\!\!\frac{\bbf^{\HH}\bB(\rho)\bbf}{\bbf^{\HH}\bC(\rho)\bbf+\bbf^{\HH}\bE\bbf+\sigma^2}$, and $f_2[t(\rho)]\!\!=\!\!\frac{1}{t(\rho)}$, in which $t(\rho)\!\!:=1\!+\!(1-\rho)P_s\bh_{sr}^{\HH}\bA^{-1}(\rho)\bh_{sr}$. And the optimal value of objective function with given $\bar{\rho}$ is denoted as $f(\bar{\rho},\bbf_{\bar{\rho}})$. Hence, the optimal solution $\!\!(\rho^\star,\bbf^\star\!)$ to problem (\ref{eqn1:optimization problem}) can be obtained by
\begin{equation}
\label{optimal solution}
(\rho^\star,\bbf^\star)=\text{arg}\max_{\bar{\rho}\in [0,1]} f(\bar{\rho},\bbf_{\bar{\rho}}).
\end{equation}

The details of GOA is summarized in Algorithm 1 as
\begin{center}
\begin{tabular}{l@{}}
\hline
\qquad \quad \ \textbf{Algorithm 1:}\  The global optimal algorithm\\
\hline
\!\!\textbf{0}: Define $\Delta\rho=\frac{1}{M}=\epsilon$, where $M$ is a large constant.\\
\!\!\textbf{1}: For $j=1,\cdots,M$, do Step \text{S0}$\sim$\text{S2}, \\
\!\!\quad and seek for the global optimal solution $\!(\rho^\star,\bbf^\star\!)$ for (\ref{eqn1:optimization problem}). \\
\!\!\!\quad \text{S0}: Denote $\bar{\rho}=j\Delta\rho$. Solve problem \eqref{eqn1:relaxed subproblem with given rho} with the given\\
\!\!\quad \quad \ \ $\bar{\rho}$, and denote $\!(\xi_{\bar{\rho}}^\star,\bZ_{\bar{\rho}}^\star\!)$ as the optimal solution to \eqref{eqn1:relaxed subproblem with given rho};\\
\!\!\!\quad \text{S1}: Obtain $\bX_{\bar{\rho}}^\star$ by $\bX_{\bar{\rho}}^\star=\frac{\bZ_{\bar{\rho}}^\star}{\xi_{\bar{\rho}}^\star}$. Do the following\\
\!\!\quad \quad \quad procedures to obtain the rank-one solution $\bbf_{\bar{\rho}}^\star$\\
\!\!\quad \quad\textbf{If} rank($\bX_{\bar{\rho}}^\star$)=1\\
\quad \quad The optimal value of problem (\ref{eqn1:optimization problem with given rho}) is $\bbf_{\bar{\rho}}^\star$, $\bX_{\bar{\rho}}^\star\!\!=\!\!\bbf_{\bar{\rho}}^\star\bbf_{\bar{\rho}}^{\HH\star}$;\\
\!\!\quad \quad \textbf{Else}\\
\quad \quad Resort to Lemma (\ref{lemma_zhang}) to seek for a rank-one\\
\quad \quad decomposition of $\bX_{\bar{\rho}}^\star$, i.e., $\bX_{\bar{\rho}}^\star=\bz\bz^{\HH}$ and $\bbf_{\bar{\rho}}^\star=\bz$,\\
\quad \quad which satisfies the equalities of (\ref{relaxed constraint 1}) and (\ref{relaxed constraint 2}).\\
\quad \quad $\bbf_{\bar{\rho}}^\star$ is the optimal solution with given $\bar{\rho}$.\\
\!\!\quad \quad \textbf{End If}\\
\!\!\!\quad \text{S2}: Remark $\bbf_{\bar{\rho}}^\star$ and $f(\bar{\rho},\bbf_{\bar{\rho}}^\star)$ as the optimal solution\\
\!\!\quad \quad \ \ and objective value for given $\bar{\rho}$, respectively. \\
\!\!\textbf{2}: Choose the optimal solution from the following equation\\
\!\!\quad $(\rho^\star,\bbf^\star)=\text{arg}\max_{\bar{\rho}\in [0,1]} f(\bar{\rho},\bbf_{\bar{\rho}}).$\\
  \hline
\end{tabular}
\end{center}
where $\epsilon$ is the given tolerant error.
\subsection{Local Optimal Solution}
Although the global optimal solution $(\bbf^\star,\rho^\star)$ can be found as described in Section \ref{GOS}, the high computation time of one dimension search in $\rho$ is expensive. Hence, we want to seek for some other algorithms which can better balance the performance of system and the cost of computation. In this section, a well-designed two-variable block coordinate decent (BCD) method is introduced to solve problem (\ref{eqn1:optimization problem}) converging to a local optimal solution, which is proved in \cite[Theorem 4.1]{Tseng2001}.

As it is mentioned above, $\sigma_d^2=\sigma_r^2=\sigma_c^2=\sigma^2$, then
\begin{equation}
\bA^{-1}(\rho)=\left[(1-\rho)P_d\bh_{dr}\bh_{dr}^{\HH}+(2-\rho)\sigma^2\bI\right]^{-1}.\nonumber
\end{equation}
In order to get more insight of $\rho$, we make some transformation of term $\bA^{-1}(\rho)$ to present it as the function of $\rho$ instead of matrix formulation, like
\begin{equation}\label{A tranformation}
\!\bA^{-1}\!(\rho)\!=\!\!\frac{\bI+\frac{1-\rho}{(2-\rho)\sigma^2}P_d\bh_{dr}^{\HH}\bh_{dr}\bI\!-\!\frac{1-\rho}{(2-\rho)\sigma^2}P_d\bh_{dr}\bh_{dr}^{\HH}}{(2-\rho)\sigma^2+(1-\rho)P_d\bh_{dr}^{\HH}\bh_{dr}},
\end{equation}
the details of this procedure will be shown in Appendix \ref{B}.

It is noted that different from the global optimal algorithm, searching $\rho^\star$ to yield the maximal secrecy rate in the interval $[0,1]$, the local optimal search for $\rho$ implies that we should select $\rho$ making the secrecy rate greater than zero to guarantee the secure source-destination transmission. Notice that the choice of $\rho$ decides the nature of secure transmission in this relay networks. For example, small $\rho$ means that there is more information eavesdropped by the untrusted relay (see \eqref{equ:information rate of relay}), and less harvested energy for the second-hop transmission, resulting in lower achievable rate for D. Meanwhile, although the large $\rho$ (approaches to 1) leads to more harvested energy and less information leakage, the confidential information is weakened and difficult to be distinguished from the noise at the destination.

Therefore, it is important to choose a suitable $\rho$ for this relay networks. In this subsection, we resort to BCD to update $\rho$ and $\bbf$ in sequence. Due to the property of BCD, the initial point has a great effect on the solution of problem (\ref{eqn1:optimization problem}). Hence, we choose multiple initial points, such as $\rho^j$ and $\bbf^j, j=1,\cdots,J$, to achieve different objective values, and the maximal one is set to be the local optimal value.

For each initial point ($\rho$, $\bbf$) (denote $\rho=\rho^j$ and $\bbf=\bbf^j,\forall j$ for simplicity), problem (\ref{eqn1:optimization problem}) with a given fixed $\bbf_k$ to achieve the temporary optimal solution $\rho_k^\star$ in the $k$-th iteration, can be written as
\begin{eqnarray}
\label{subeqn:subproblem with given f}
\max_{\rho} &&\scriptsize{\left(\!1\!+\!\frac{\bbf_k^{\HH}\bB(\rho)\bbf_k}{\bbf_k^{\HH}\bC(\rho)\bbf_k\!+\!\bbf_k^{\HH}\bE\bbf_k\!+\!\sigma^2}\!\right)\!\!\left(\!\!\frac{1}{1\!+\!(1-\rho)P_s\bh_{sr}^{\HH}\bA^{-1}(\rho)\bh_{sr}}\!\!\right)}\nonumber\\
\text{s.t.} &&\bbf_k^{\HH}\bT(\rho)\bbf_k\leq P_r^{\text{EH}}.
\end{eqnarray}

Plugging (\ref{A tranformation}) in (\ref{subeqn:subproblem with given f}), it is easy to verify that the objective function of problem (\ref{subeqn:subproblem with given f}) can be rewritten as an univariate function of scaler $\rho$ with high order term, and the constraint with respect to $\rho$ is convex. Therefore, the optimal solution $\rho_k^\star$ of the above problem can be obtained. We check the monotonicity of $f(\rho,\bbf_k)$ with respect to $\rho$ (the details are given in Appendix \ref{C}), choose the points where the derivatives of $f(\rho,\bbf_k)$ with respect to $\rho$ equal to zero, calculate the values of the objective function in (\ref{subeqn:subproblem with given f}) with those chosen points, and select the one with the maximal value. If there is no point making the derivative of $f(\rho,\bbf_k)$ with respect to $\rho$ equal to zero in the range of $[0,1]$, it means that $f(\rho,\bbf_k)$ is monotonic, and the point at the boundary of the constraint will be chosen to be the optimal one. The reason is given in the Remark \ref{remark2}.

\begin{remark}\label{remark2}
As $f$ is a smooth function, $\nabla f$ should be a continuous function. We have that if $\nabla_{\rho} f$ can not obtain $0$, $\nabla_{\rho} f$ must be positive or negative in the whole interval [0,1]. If not, let us assume that $\nabla_{\rho} f(\rho_1, \bbf_k)$ and $\nabla_{\rho} f(\rho_2, \bbf_k)$ have different signs. Because of the continuity of $\nabla_{\rho}f$, there should be a point $\rho_3\in (\rho_1,\rho_2)$ such that $\nabla_{\rho} f (\rho_3, f_k)=0$. This is a contradiction.
\end{remark}

Then we can update $\bbf$ depending on the obtained $\rho_{k+1}$ (where $\rho_{k+1}:=\rho_k^{\star}$) in the $(k+1)$-th iteration through solving the following relaxed optimization problem referred to the procedures of (\ref{eqn:relaxed problem with given rho}) to (\ref{eqn1:relaxed subproblem with given rho})
\begin{eqnarray}
\label{subeqn1:relaxed subproblem with given rho}
\max_{\bZ_{k+1},\xi} &&\Tr{\left(\bB(\rho_{k+1})\bZ_{k+1}\right)}\\
\text{s.t.} &&\Tr{\left(\bC(\rho_{k+1})\bZ_{k+1}\right)}+\Tr{\left(\bE\bZ_{k+1}\right)}+\xi\sigma^2=1,\nonumber \\
&&\Tr{\left(\bT(\rho_{k+1})\bZ_{k+1}\right)}\leq \xi P_r^{\text{EH}}, \nonumber\\
&&\bZ_{k+1}\succeq\bzero,\xi > 0, \nonumber
\end{eqnarray}
the optimal solution of which can be efficiently obtained by solvers, e.g., CVX. Denote $(\bZ_{k+1}^\star,\xi^\star)$ as the optimal solution of problem (\ref{subeqn1:relaxed subproblem with given rho}) in the $(k+1)$-th iteration, and we achieve the optimal solution of $\bX^\star$ by $\bX_{k+1}^\star=\frac{\bZ_{k+1}^\star}{\xi^\star}$. Resorting to lemma \ref{lemma_zhang}, we can get a rank-one decomposition from $\bX_{k+1}^\star=\bbf_{k+1}^\star\bbf_{k+1}^{\star\HH}$, i.e., $\bbf_{k+1}^\star$ is the optimal solution in the $(k+1)$-th iteration.

In the following, we present the procedures of the local optimal algorithm (LOA) in Algorithm 2.
\begin{center}
\begin{tabular}{l@{}}
\hline
\qquad  \ \textbf{Algorithm 2:}\  The LOA based on BCD method\\
\hline
\!\!  \textbf{0} (Initialization): Randomly generate feasible points $\!\!(\rho^j,\bbf^j\!)$,\\
\!\!\quad $\forall j$, and let the initial objective values be $f(\rho^j,\bbf^j), \forall j$.\\
\!\!\quad Do Step\textbf{1}-\textbf{2} for each $j$, denote $\rho=\rho^j$ and $\bbf=\bbf^j$\\
\!\!\quad for simplicity, and set $k=0$.\\
\!\!\textbf{1} (Block Coordinate Maximization): \\
\!\!\quad (1)\;Solve Problem (\ref{subeqn:subproblem with given f}) with $f\!(\rho,\bbf_k\!)$, and \\
\!\!  \quad \quad \ get optimal solution $\rho_{k+1}$;\\
\!\!\quad (2)\;Solve Problem (\ref{subeqn1:relaxed subproblem with given rho}) with $f(\rho_{k+1},\bbf)$, and\\
\!\!  \quad \quad \ get optimal solution $\bbf_{k+1}$ resorting to Lemma \ref{lemma_zhang};\\
\!\!\quad Set $\Delta=|f(\rho_{k+1},\bbf_{k+1})-f(\rho_k,\bbf_k)|/|f(\rho_{k+1},\bbf_{k+1})|$.\\
\!\!  \textbf{2} (Stopping Criterion): If $\max(0,\Delta) \le \epsilon$, stop and return\\
\!\!\quad  $(\rho_k,\bbf_k)$; otherwise, set $k:=k+1$ and go to Step 1.\\
\!\!  \textbf{3} (Local Optimal Solution): Denote $(\rho^j_k,\bbf^j_k)$ and $f(\rho^j_k,\bbf^j_k)$\\
\!\!\quad as the solution and objective value of problem (\ref{eqn1:optimization problem}) with\\
\!\!\quad the given initial point $(\rho^j,\bbf^j)$, hence the local optimal\\
\!\!\quad solution is chosen by\\
\!\!\quad \quad$(\rho^\star,\bbf^\star)=\text{arg}\max_j f(\rho^j_k,\bbf^j_k).$\\
  \hline
\end{tabular}
\end{center}

\section{Simulation Results}
In this section, we investigate the performance of the proposed algorithms via numerical simulations. The path loss model for the energy harvesting relay channel is Rayleigh distributed and denoted by $|\beta|^2d^{-2}$, where $|\beta|$ and $d$ represent the short-term channel fading and the distance between two nodes (S and R, D and R, R and D). $|\beta|^2$ follows the exponential distribution with unit mean. We set the noise powers as $\sigma_{r}^2\!=\!\sigma_d^2\!=\!\sigma_c^2\!=\!\sigma^2\!=\!0\text{dBm},\ \forall i$, the number of antennas at R as 2, and the energy conversion efficiency $\eta=1$.

\begin{figure}
  \centering
  \includegraphics[height=55mm, origin=br]{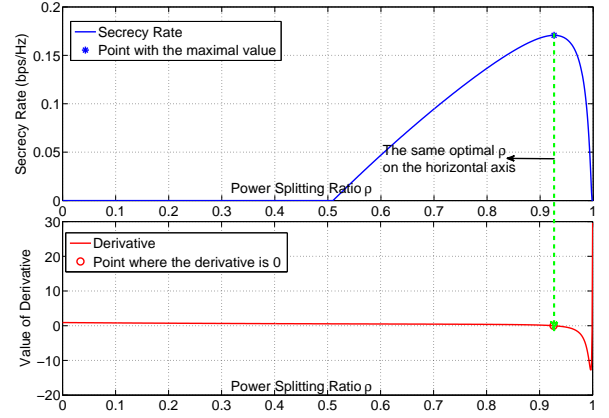}\\
   \caption{Verification of Remark \ref{remark1} that the optimal $\rho^\star$ is obtained when the derivative $\frac{d R_{sr}}{d\rho}(\rho^\star)$ equals zero, $P_s=P_d=40\text{dBm}$.}\label{2}
\end{figure}

To get more insight of Remark \ref{remark1}, we verify the conclusion of Remark \ref{remark1} using one channel realization. For the single-antenna untrusted relay case, the optimal PS ratio $\rho^\star$ can be achieved by setting $\frac{d R_{sr}}{d\rho}=0$. As shown in Fig.\ref{2}, the point with the maximal secrecy rate in the upper subfigure corresponds to the point $\rho^\star$ where $\frac{d R_{sr}}{d\rho}(\rho^\star)=0$ on the horizontal axis in the second subfigure. In addition, we conclude that both small $\rho$ and large $\rho$ (approaches to\! 1\!) yield smaller secrecy rate.

\begin{figure}
  \centering
  \includegraphics[height=55mm, origin=br]{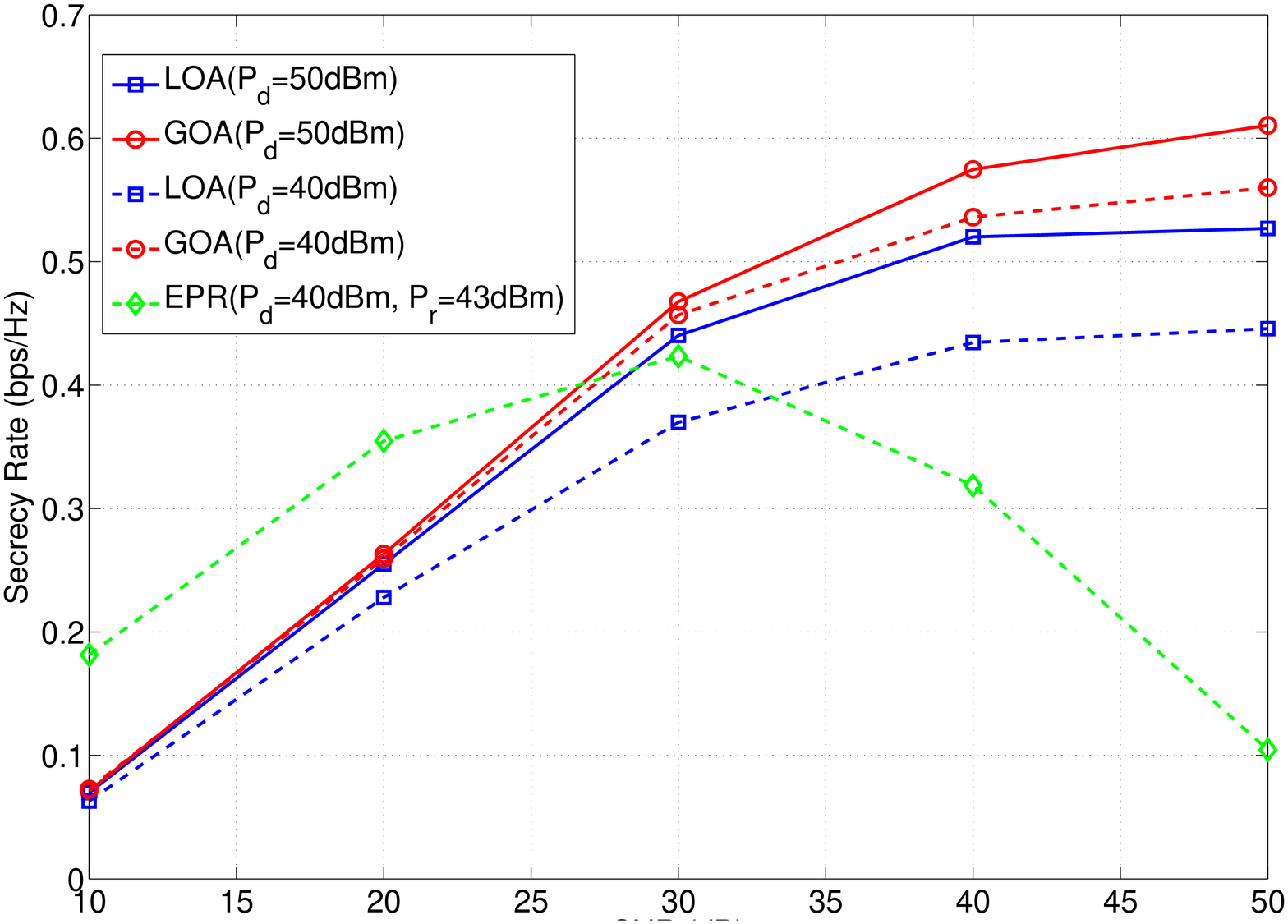}\\
   \caption{Comparison of the secrecy rate achieved by different algorithms, which is averaged over 100 channel realizations, and SNR=$\frac{P_s}{\sigma^2}$.}\label{3}
  \includegraphics[height=55mm, origin=br]{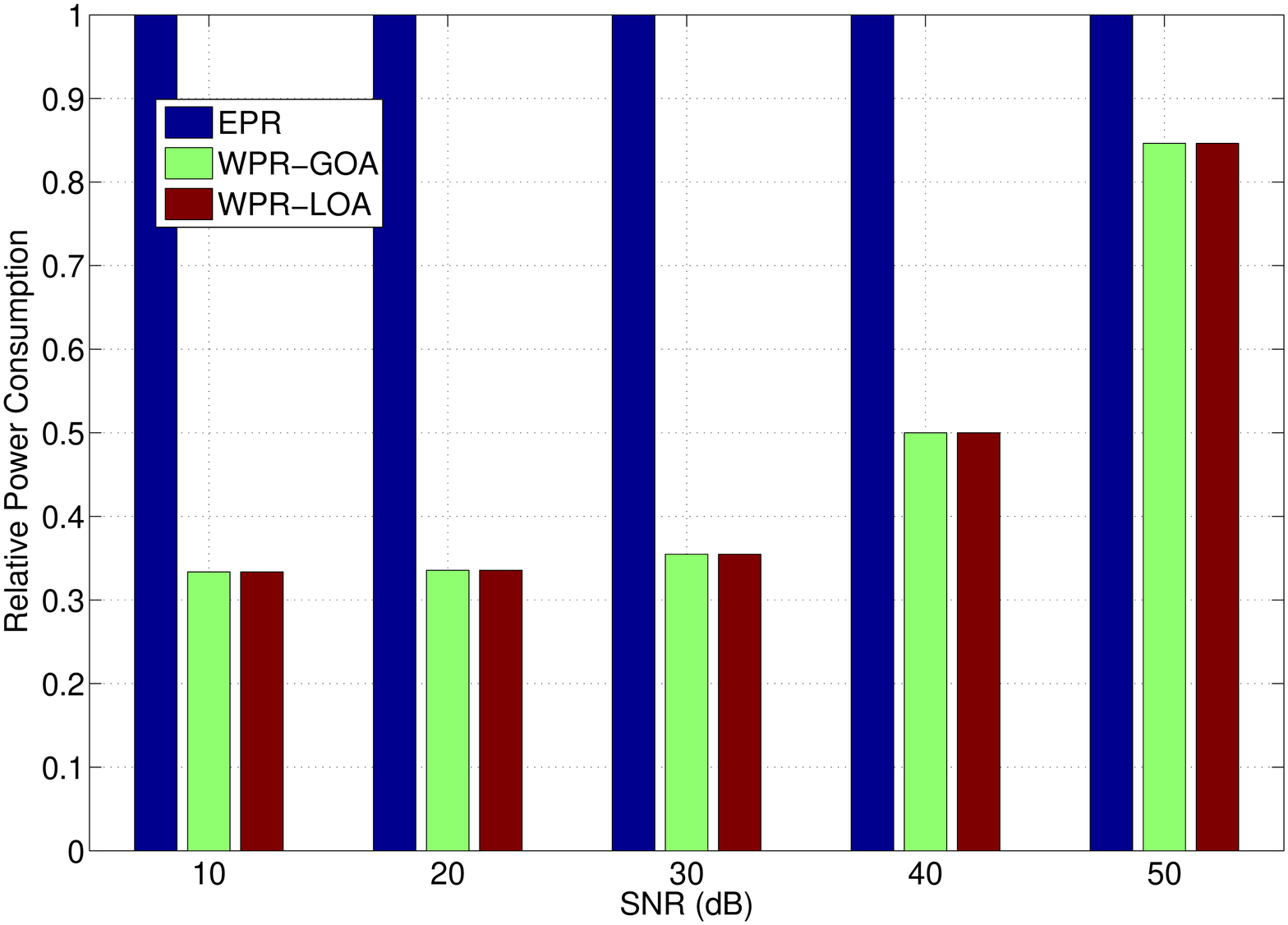}\\
   \caption{Comparison of relative power used (relative to the power consumption of EPR case with $P_d=40\text{dBm}$ and $P_r=43\text{dBm}$.}\label{4}
\end{figure}

The secrecy rate performances for GOA and LOA in MIMO untrusted relay networks are given in Fig.\ref{3}. In the low SNR region, the performance achieved by LOA quite approximates the one in GOA. The reason is that the key element affecting the performance in that SNR region is the harvested energy powering the next-hop transmission. Although we can get the optimal $\rho^\star$ resorting to GOA, it just helps a little.

In contrast, the dominant factor for the performance is the power splitting ratio $\rho$ in the high SNR region. It is the fact that the higher the SNR, the larger gap between GOA and LOA for the fixed AN power in Fig.\ref{3}. Interestingly, as the power of AN ($P_d$) increases, the performance of LOA is getting much closer to that of GOA for the whole region of SNR. For instance, at SNR=50dBm, the relative performance ratio is growing from 79.6\% to 86.3\% as $P_d$ raises from 40dBm to 50dBm.

The comparison of computation time for GOA and LOA is shown in Table.\ref{table1} with different given $\epsilon$. The results are averaged among 100 channel realizations, and the unit of measurement for computation time is \emph{second}(s). It is obvious that LOA takes much less computation time than GOA. We can figure out that LOA is relatively efficient, and it can achieve satisfactory secrecy rate performance compared to GOA, especially when $P_d$ is large, but with less computation time.

\begin{table}
\caption{Comparison of The Computation Time in GOA and LOA for per channel realization.}\label{table1}
\begin{tabular}{|c|c|c|c|c|}
\hline
\multirow{2}{*}{Method} & \multicolumn{3}{|c|}{$\epsilon$} \\
\cline{2-4}
& $10^{-2}$ & $10^{-3}$ & $10^{-4}$\\
\hline
Local Optimal Algorithm (LOA)&\ \ \bf{0.2896} &\ \ \bf{2.9868} &\ \ \bf{7.4535} \\
\hline
Global Optimal Algorithm (GOA) &\ \ 0.9706 &\ \ 6.1656 &\ \ 47.9588 \\
\hline
\end{tabular}
\end{table}

To find more knowledge of the benefits of the wireless-powered relay, we compare the performances between the wireless-powered relay (WPR) and the external-powered relay (EPR, i.e., the power of the relay is supplied by the power grid or battery, rather than harvested from its received signal) in Fig.\ref{3}. We also show the relative power consumption of the three algorithm (GOA, LOA and EPR) in Fig.\ref{4}, and denote GOA and LOA as WPR-GOA and WPR-LOA, respectively. It is clear that WPR consumes less power than EPR because the power of WPR is supplied by energy harvesting from the received signal, rather than the external power for EPR case.

The secrecy rate of EPR with the given constant relay power $P_r=43\text{dBm}$ is almost twice times than those of GOA and LOA in the low region of SNR. Because WPR can not harvest enough energy to achieve better performance in that low SNR region compared with EPR. By contrast, in the high SNR region, WPR can harvest sufficient energy to support its next-hop transmission and obtain higher secrecy rate performance with reasonable PS ratio for the information processing, but consumes less energy compared with EPR in Fig.\ref{4}. As for EPR, more information is eavesdropped by the untrusted relay as the source power $P_s$ is increasing in the high SNR region, hence, its secrecy rate performance is reducing. From the aspect of secrecy rate performance, WPR has advantage over EPR for the reason that it can harvest energy from the wireless signal and adjust the PS ratio to achieve better performance with less information leakage and power consumption, especially in the high SNR region.

\section{Conclusion}
In this paper, we study the secrecy rate maximization problem in the wireless-powered untrusted relay networks according to joint power splitting and secure beamforming design. We propose two algorithm named as Global optimal algorithm (GOA) and Local optimal algorithm (LOA) with well established convergence result to jointly optimize the PS ratio and relay beamforming, and LOA can achieve satisfactory secrecy rate performance compared to GOA, but with less computation time. We also show that the wireless-powered untrusted relay has advantage over external-powered untrusted relay at the aspect of secrecy rate performance but with less power consumption, especially in the high SNR region.

\appendices
\section{Transformation of $\bA^{-1}(\rho)$}
\label{B}
We rewrite $\bA^{-1}(\rho)$ to represent it as the function of $\rho$ instead of matrix formulation, as follows
\begin{eqnarray}
\!\!\!\!\bA^{-1}(\rho)\!\!\!\!&=&\!\!\!\!\left[(2-\rho)\sigma^2\bI+(1-\rho)P_d\bh_{dr}\bh_{dr}^{\HH}\right]^{-1}\nonumber\\
\!\!\!\!&=&\!\!\!\!\frac{1}{(2-\rho)\sigma^2}\bI\!\!-\!\!\frac{\frac{1}{(2-\rho)\sigma^2}\bI\left[\!(\!1\!-\!\rho\!)P_d\bh_{dr}\bh_{dr}^{\HH}\right]\frac{1}{(2-\rho)\sigma^2}\bI}{1+(1-\rho)P_d\bh_{dr}^{\HH}\frac{1}{(2-\rho)\sigma^2}\bI\bh_{dr}}\nonumber\\
\!\!\!\!&=&\!\!\!\!\frac{\bI+\frac{1-\rho}{(2-\rho)\sigma^2}P_d\bh_{dr}^{\HH}\bh_{dr}\bI-\frac{1-\rho}{(2-\rho)\sigma^2}P_d\bh_{dr}\bh_{dr}^{\HH}}{(2-\rho)\sigma^2+(1-\rho)P_d\bh_{dr}^{\HH}\bh_{dr}},
\end{eqnarray}
where the second equation comes form the definition of \emph{Sherman-Morrison-Woodbury} formula\cite{MatrixAnalysis}, which is
\begin{equation}
\label{Sherman-Morrison-Woodbury formula}
\left[\bK+\bx\by^{\HH}\right]^{-1}=\bK^{-1}-\frac{\bK^{-1}\bx\by^{\HH}\bK^{-1}}{1+\by^{\HH}\bK^{-1}\bx}.
\end{equation}

\section{The Derivative of $f(\rho,\baf)$ with respect to $\rho$}
\label{C}
Recall the those denotations in Section \ref{GOS} of $f(\rho,\bbf)$, $f_1(\rho,\bbf)$ and $f_2[t(\rho)]$, we have the partial derivative of $f_1$ with respect to $\rho$ like
\begin{eqnarray}
\!\!\!\!\frac{\partial f_1(\rho,\baf)}{\partial \rho}\!\!\!\!\!\!&=&\!\!\!\!\!\!\frac{-\!\!P_s\baf^{\HH}\!\!\left[\!\!\left(\bh_{sr}\bh_{sr}^{\HH}\right)^T\!\!\!\!\!\!\otimes\!\!\left(\bh_{rd}\bh_{rd}^{\HH}\right)\!\!\right]\!\!\baf\!\!\left(\baf^{\HH}\bE\baf\!\!+\!\!\sigma^2\!\!\right)}{\left(\baf^{\HH}\bC(\rho)\baf+\baf^{\HH}\bE\baf+\sigma^2\right)^2}.
\end{eqnarray}
And the derivative of $f_2$ with respect to $\rho$, is given
\begin{eqnarray}
&&\frac{\partial f_2[t(\rho)]}{\partial \rho}=-\frac{1}{t(\rho)^2}\frac{\partial t(\rho)}{\partial \rho}\nonumber\\
&&=-\frac{1}{t(\rho)^2}\left[\frac{-P_s\sigma^2}{\left((2-\rho)\sigma^2+(1-\rho)P_d\bh_{dr}^{\HH}\bh_{dr}\right)^2}m+\nonumber\right.\\
&&\left.\frac{P_d\bh_{sr}^{\HH}\bh_{dr}\bh_{dr}^{\HH}\bh_{sr}-P_d\bh_{dr}^{\HH}\bh_{dr}\bh_{sr}^{\HH}\bh_{sr}}{(2-\rho)^2\sigma^2}n\right],\nonumber
\end{eqnarray}
where $t(\rho)$, taking account of the new transformation of $\bA^{-1}(\rho)$, is written as
\begin{equation}
t(\rho)=1+mn,\nonumber
\end{equation}
where $m=$
\begin{equation}
\!\!\bh_{sr}^{\HH}\bh_{sr}\!\!+\!\!\frac{\!1\!-\!\rho}{(\!2\!-\!\rho)\sigma^2}\!P_d\!\bh_{dr}^{\HH}\bh_{dr}\bh_{sr}^{\HH}\bh_{sr}\!\!-\!\!\frac{\!1\!-\!\rho}{\!(\!2\!-\!\rho\!)\sigma^2}\!P_d\bh_{sr}^{\HH}\bh_{dr}\bh_{dr}^{\HH}\bh_{sr}\nonumber,
\end{equation}
and $n=\frac{(1-\rho)P_s}{(2-\rho)\sigma^2+(1-\rho)P_d\bh_{dr}^{\HH}\bh_{dr}}$.

The derivative of $f(\rho,\bbf_k)$ with respect to $\rho$ is given by
\begin{eqnarray}
\!\!\frac{\partial f(\rho^{\star},\baf)}{\partial \rho^{\star}}\!\!\!\!\!\!&=&\!\!\!\!\!\!\frac{\partial f_1(\rho^{\star},\baf)}{\partial \rho^{\star}}f_2[t(\rho^{\star})]\!\!+\!\!\frac{\partial f_2[t(\rho^{\star})]}{\partial \rho^{\star}}f_1(\rho^{\star},\baf).
\end{eqnarray}

\bibliographystyle{IEEEbib}
\bibliography{reference}

\end{document}